\documentclass[conference,10pt]{IEEEtran}

\IEEEoverridecommandlockouts

\usepackage{hyphenat}
\usepackage{cite}
\usepackage{amsmath,amssymb,amsfonts}
\usepackage{amsthm}

\usepackage{graphicx}
\usepackage{textcomp}
\usepackage{xcolor}
\usepackage[inline]{enumitem}
\usepackage{booktabs}
\usepackage{array}
\usepackage{multirow}
\usepackage{multirow}
\usepackage{hyphenat}
\usepackage{booktabs}
\usepackage{tabularx}
\usepackage{hyperref}
\usepackage{balance}
\usepackage{acronym}
\usepackage{makecell}
\hypersetup{colorlinks=true, linkcolor=blue, citecolor=blue, urlcolor=blue}
\usepackage{fancyref}
\usepackage{xspace}
\usepackage[final]{listings}
\usepackage{todonotes}
\usepackage{comment}
\newcommand{\eg}{\emph{e.g.,}\xspace}

\acrodef{RL}{Reinforcement Learning}
\acrodef{AI}{Artificial Intelligence}
\acrodef{GAI}[GenAI]{Generative AI}
\acrodef{LLM}{Large Language Model}
\acrodef{AIOps}{AI-Driven IT Operations}

{\acresetall
   \acused{AI}
   }

\newcommand{\linebreakand}{%
  \end{@IEEEauthorhalign}
  \hfill\mbox{}\par
  \mbox{}\hfill\begin{@IEEEauthorhalign}
}
\makeatother

\begin{document}

\title{Mitigating Uncertainty Interactions in GenAI-based Adaptive Systems: Vision, Challenges and Preliminary Guidelines}
%\title{Towards a Multi-Level Framework for Uncertainty Mitigation in\\Generative AI-Based Software Intensive Systems}
%catalogue of strategies for mitigation...

\author{
\IEEEauthorblockN{Karthik Vaidhyanathan}
\IEEEauthorblockA{\textit{SERC, IIIT Hyderabad}\\India}
\and
\IEEEauthorblockN{Norha Villegas}
\IEEEauthorblockA{\textit{Universidad Icesi}, Colombia\\\textit{University of Victoria}, Canada}
\and
\IEEEauthorblockN{Raffaela Mirandola}
\IEEEauthorblockA{\textit{Karlsruhe Institute of}\\\textit{Technology (KIT)}, Germany}
\and
\IEEEauthorblockN{Nicol\'{a}s Cardozo}
\IEEEauthorblockA{\textit{Universidad de Los Andes}\\Colombia}
\and
\IEEEauthorblockN{Ivana Dusparic}
\IEEEauthorblockA{\textit{Trinity College Dublin}\\Ireland}
\and
\IEEEauthorblockN{Marin Litoiu}
\IEEEauthorblockA{\textit{York University}\\Canada}
\and
\IEEEauthorblockN{Ralf Reussner}
\IEEEauthorblockA{\textit{Karlsruhe Institute of}\\\textit{Technology (KIT)}, Germany}
\and
\IEEEauthorblockN{Gabriel Tamura}
\IEEEauthorblockA{\textit{Universidad Icesi}, Colombia\\\textit{University of Victoria}, Canada}
}

%\IEEEpeerreviewmaketitle
\maketitle

\begin{abstract}
Modern software-intensive systems increasingly incorporate \acf{GAI} components, including \acp{LLM} and agentic subsystems, which introduce novel and compounding sources of uncertainty across system layers.
The self-adaptive systems community has made significant strides in understanding and managing uncertainty. The intrinsic characteristics of \ac{GAI}, including probabilistic outputs, hallucinations, context window limitations, and memory staleness, demand a re-examination of existing frameworks and mitigation strategies for dealing with uncertainty, and especially, the \emph{interactions} among uncertainty occurrences and manifestations.
This position paper posits an initial conceptual framework that establishes preliminary guidelines for characterizing and mitigating \emph{uncertainty interactions} in \ac{GAI}-based software-intensive systems.
We argue that mitigation must be considered across the full software lifecycle, encompassing requirements analysis, design-time decisions, and run-time adaptations, which implies the unfeasibility of decoupling it from monitoring and analysis at execution.
We introduce the notion of a \emph{mitigation function} that maps uncertainty interaction categories, involved uncertainty types, system design characteristics, criticality, and mitigating architectural \emph{tactics} to impacted quality attributes and corresponding tradeoffs.
As a proof of concept, we apply the framework to three illustrative use cases: LLM serving, AI-driven software operations, and agentic cloud management.
\end{abstract}


%% ============================================================
% !TEX root = main.tex

\section{Introduction}
\label{sec:intro}

Software intensive systems are increasingly augmented with \ac{GAI} capabilities.
\acp{LLM} are deployed not only as conversational interfaces but as core reasoning and decision-making components in autonomous systems~\cite{xi2023rise}.
Subsystems with multi-agent architectures, in which several LLM-based agents collaborate to achieve a common goal, are being used for various tasks such as scientific discovery~\cite{boiko2023autonomous} and
cloud operations~\cite{parthasarathy2025cloudops}. This shift introduces new uncertainty forms that interact in ways the existing self-adaptive systems literature has not fully addressed~\cite{hezavehi2021uncertainty,weyns2023b}.

Over the years, the self-adaptive systems community has developed mature frameworks for reasoning about uncertainty, its sources, nature, interactions, and mitigation~\cite{esfahani2013uncertainty,mahdavi2017classification,troya2021uncertainty}. Moreover, 
the community has also established a shared vocabulary for characterizing uncertainty across problem and solution spaces~\cite{weyns2023b}.
However, \ac{GAI} components challenge these frameworks in fundamental ways. Unlike traditional software components, where uncertainty is largely external, \ac{GAI}-based systems introduce both \emph{aleatoric} and \emph{epistemic} uncertainties that are intrinsic to system behavior, and even from its development. Aleatoric uncertainty arises from variability in inputs and execution (e.g., workload fluctuations, stochastic decoding, re-training processes), while epistemic uncertainty stems from incomplete or imperfect knowledge (e.g., hallucinations, model limitations, or stale memory). In GenAI-based and agentic systems, these uncertainties do not occur in isolation but \emph{interact and propagate across components} (e.g., models, memory, and tools), leading to emergent behaviors that are difficult to reason about using traditional approaches. This position paper outlines a vision for characterizing and mitigating uncertainty interactions in \ac{GAI}-based software-intensive systems. We argue that the interacting and compounding nature of uncertainty in these systems requires new abstractions that connect uncertainty reasoning with architectural decision-making. Thus, we introduce the mitigation function as a central abstraction for reasoning about the impact of uncertainty interactions on system quality attributes, making the following contributions: 
\begin{enumerate*}[label=(\arabic*)]
\item The definition of the \emph{mitigation function} as a unifying abstraction that maps uncertainty interactions, system context, and criticality to quality attribute outcomes through architectural tactics, enabling systematic reasoning about trade-offs; 
\item the instantiation of the proposed abstraction across three representative use cases illustrating its applicability and generality; and 
\item a vision and research agenda for systematic uncertainty mitigation in Gen-AI systems, identifying key challenges such as uncertainty interaction characterization, mitigation strategy selection, cross-layer propagation, trade-off evaluation, and dynamic context management. The proposed agenda provides a roadmap for further research.
\end{enumerate*}

%We ground this framework in three use cases drawn from ongoing research and industrial collaborations:
%\begin{enumerate}[leftmargin=0.5cm]
%    \item \textbf{\ac{LLM} Serving under SLO Constraints:} Managing tail latency in \ac{LLM} inference systems facing workload and output uncertainty.
%    \item \textbf{\ac{AIOps}:} Using Software Twins as digital replicas of software applications that combine proactive risk assessment of operational changes and root-cause analysis of runtime failures to prevent them. \cite{rivera2023aiops-risk}
%    \item \textbf{Agentic \ac{AI} for Cloud Operations:} Autonomous multi-agent systems managing cloud infrastructure through automated workflows.

%% ============================================================
\section{Background and Motivation}
\label{sec:background}

\subsection{Uncertainty in Self-Adaptive Systems}

Uncertainty can be classified along several dimensions: its \emph{location} (where in the system it arises), its \emph{nature} (epistemic vs.\ aleatory), its \emph{level} (statistical, scenario-based, or recognized ignorance), and its \emph{emerging time} (requirements, design, or runtime)~\cite{hezavehi2021uncertainty,ramirez2012taxonomy}.
Uncertainty \emph{interactions} occur when multiple uncertainty sources combine, potentially through independence, dominance, augmentation, or conflict~\cite{troya2021uncertainty}.
The interaction category is central to understanding how uncertainties interact: dominance means one source subsumes another, augmentation means sources amplify each other, and conflict means that uncertainties generate contradictory impacts on the system.

%\subsection{Quality Attributes and Architectural Tactics}

%In software architecture, a \emph{quality attribute} is a measurable or testable property of a system used to indicate how well the system satisfies stakeholder needs beyond its bare functionalities~\cite{bass2021software}.
%    Examples include latency, availability, safety, security, and modifiability.
%An \emph{architectural tactic} is a design decision that directly influences the achievement of a quality attribute response.
%Tactics are the fundamental building blocks from which architectural patterns and styles are composed.
%For instance, tactics for achieving performance include ``task parallelization and distribution,'', ``manage sampling rate,'' ``limit event response,'' and ``prioritize events''~\cite{bass2021software}.
%The key insight we bring to the uncertainty mitigation problem is that \textbf{tactics serve as the mechanism through which uncertainty is addressed, and quality attributes are measurable outcomes of applying those tactics in the presence of specific uncertainty interactions}.
%Our claim is that a tactic is not chosen in isolation; it is chosen \emph{because of} a specific uncertainty interaction, and its effectiveness is measured through the quality attributes it influences.

\subsection{Uncertainty in \ac{GAI} Systems}

\ac{GAI} introduces several distinctive uncertainty sources that go beyond classical software uncertainty:

\begin{itemize}[nosep]
    \item \textbf{Non-deterministic output:} Even with identical inputs, LLMs produce varying outputs due to stochastic decoding (temperature, top-$k$, nucleus sampling), making output quality and length inherently uncertain.
    \item \textbf{Hallucination:} LLMs can generate plausible but factually incorrect content, introducing epistemic uncertainty that is difficult to detect without external grounding.
    \item \textbf{Context window limitations:} Finite context windows create structural uncertainty about whether all relevant information is available for reasoning.
    \item \textbf{Memory staleness:} In agentic systems with persistent memory, retrieved context may be modified, outdated, incomplete, or irrelevant, compounding reasoning uncertainty.
    \item \textbf{Tool orchestration variability:} Agents selecting and parameterizing external tools introduce uncertainty in both the selection and execution dimensions.
    \item \textbf{Multi-agent coordination:} When multiple agents collaborate, uncertainties propagate and interact across agent boundaries, often amplifying in ways not observable within any single agent.
\end{itemize}

Further, the uncertainties can also stem from the environment in which these systems operate. These sources interact in complex ways. For instance, a hallucinated intermediate result from one agent may be stored in shared memory, retrieved by another agent, and used to parameterize a tool function, creating a cascade of correlated and accumulated imprecise results that no single-agent analysis would reveal.
Crucially, the \emph{interaction category} between these sources (\eg whether hallucination \emph{dominates} tool selection, or whether memory staleness and output non-determinism \emph{augment} each other) determines the appropriate mitigation response.

%% ============================================================
\section{The Uncertainty Mitigation Framework}

\begin{figure*}[tphb]
    \centering
    \includegraphics[width=0.7\textwidth]{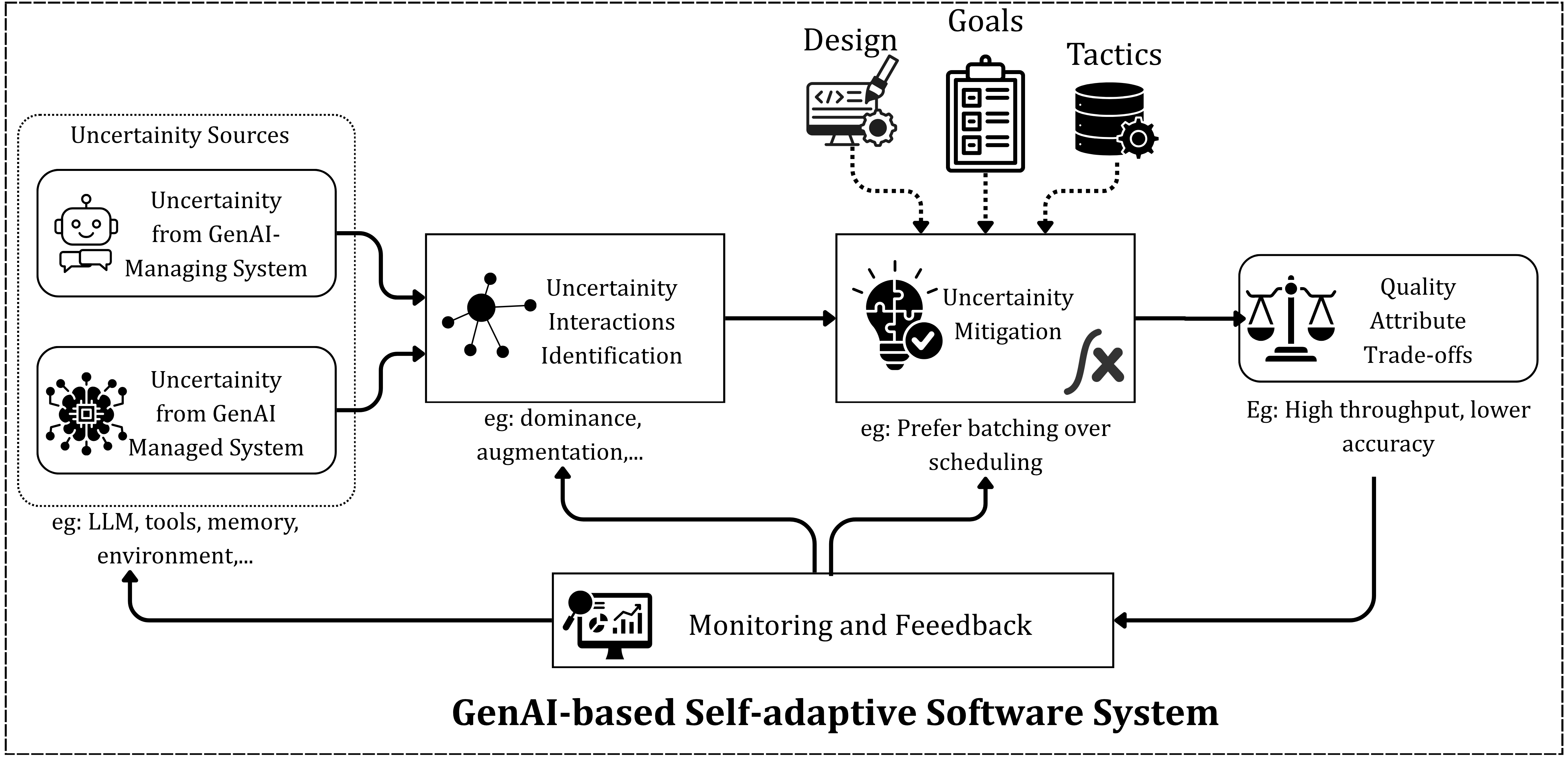}
    \caption{Uncertainty Mitigation Framework for GenAI-based Self-adaptive Systems}
    \label{fig:mitigation}
\end{figure*}

Figure \ref{fig:mitigation} provides a high-level view of the proposed mitigation framework for GenAI-based self-adaptive systems. The process begins by identifying uncertainty sources originating from both the managing (adaptive) system and the managed system. These uncertainties are then analyzed to characterize their interactions (\eg dominance, augmentation, conflicting). Based on this characterization, the mitigation function leverages system design context, goals, and candidate architectural tactics to determine appropriate mitigation strategies. The outcome of this process is a set of quality attribute trade-offs (\eg performance vs. accuracy) and corresponding tactics, which inform architectural decision-making. A monitoring and feedback loop continuously observes system behavior and outputs, enabling dynamic refinement of uncertainty characterization and mitigation decisions over time. %The proposed mitigation function is formalized as follows.
\label{sec:formalization}

%We propose a formalization for the mitigation function that connects uncertainty interactions, system design, architectural tactics, and quality attribute outcomes.
%The formalization preserves the structure established by the UNISON community for characterizing uncertainty interactions, and extends it with tactics as a new input dimension.

%\subsection{Existing Formalization: Uncertainty Interaction to Quality Attributes}

%The existing formalization from the UNISON initiative characterizes a mitigation input as a structured tuple and maps it to a quality attribute output.

We propose the following formalization for the mitigation function. The input captures:

\begin{itemize}
    \item \textbf{IC} (Interaction Category). The type of uncertainty interaction: dominance, augmentation, conflict, or independence;
    \item \textbf{Types}: The nature of the uncertainty sources involved: epistemic, aleatory, or both;
    \item \textbf{Design}: System design characteristics, including the design-time decisions (DT) as exposed in its architecture and context variables (CV) that influence the system behavior;
    \item \textbf{Cr}: The criticality vector, identifying which uncertainty sources ($U_1, U_2, \ldots$) are most critical;
    \item \textbf{Q}: Vector of quality attributes under risk;
    \item \textbf{Tactics}: Vector of \textbf{Tactic} candidates, each of which could be used to mitigate the specified types of uncertainties;
\end{itemize}

\noindent where a \textbf{Tactic} is a tuple specifying:
\begin{itemize}
    \item \textit{mechanism}: the concrete intervention (\eg admission control, what-if simulation, policy-bounded autonomy) to introduce;
    \item \textit{scope}: the architectural level ($\in \{$component, subsystem, system, ecosystem$\}$) at which the intervention applies; and
    \item \textit{lifecycle}: the moment when the mechanism is applied ($\in \{$design-time, deployment-time, runtime$\}$).
\end{itemize}

\noindent The output is a tuple specifying which quality attribute is assured, for a corresponding tactic is introduced. 
More formally:

\begin{equation}
\label{eq:extended}
\begin{split}
\textbf{Input:}\; & \big(\,\textit{IC}\!: \langle\textit{cat}\rangle,\;\textit{Types}\!: [\textit{t}_1,\; \textit{t}_2, \ldots ], \\
                  & \;\;\, \textit{Design}\!: \langle\textit{DT}, \textit{CV}\!: \ldots\rangle,\;\textit{Cr}\!: [U_i, \ldots], 
                  \;\, 
                  \mathbf{Q}\!: [q_1,\ldots, q_j], \\
                  & \;\;\, 
                  \textbf{Tactics}\!: [ Tac_{1\ldots k}:\langle\textit{mech.}, \textit{scope}, \textit{lifecycle}\rangle\ ]\big) \\[4pt]
\textbf{Output:}\; & \langle \mathbf{Q}_{\{1\ldots j\}},Tac_{\{1\ldots k \}}\rangle
\end{split}
\end{equation}
where the domains of $i,j,k$ are the cardinalities of uncertainty sources, quality attributes, and tactics, respectively. 

\smallskip
%\noindent\hspace{1em}%
%\texttt{Input:}\; $\big($%
%IC: $\langle$Dominance$\rangle$,\;
%Types: $\langle$Epistemic, Aleatory$\rangle$,

%\noindent\hspace{4.2em}%
%Design: $\langle$DT, CV: %\ldots$\rangle$,\;
%Cr: $[U_1]\big)$

%\noindent\hspace{1em}%
%\texttt{Output:}\; %$\mathbf{Q}\langle$functional %correctness$\rangle$

%\begin{equation}
%\label{eq:existing}
%\boxed{
%\begin{aligned}
%\textbf{Input:}\; & \big(\,\textit{IC}\!: \langle\textit{cat}\rangle,\;\textit{Types}\!: \langle\textit{t}_1, \textit{t}_2, \ldots\rangle, \\
%                  & \;\;\, \textit{Design}\!: \langle\textit{DT}, \textit{CV}\!: \ldots\rangle,\;\textit{Cr}\!: [U_i, \ldots]\,\big) \\[4pt]
%\textbf{Output:}\; & \mathbf{Q}\langle q_1, q_2, \ldots \rangle
%\end{aligned}
%}
%\end{equation}

\noindent In this formalization, the uncertainty interaction category ($IC$) is the \emph{key structuring concept}: it determines how the uncertainty sources combine and therefore what kind of mitigation is needed.
%\smallskip
%\noindent\textbf{Example.} Consider an uncertainty interaction where workload uncertainty \emph{dominates} output uncertainty in an LLM serving system:\todo{not sure what is intended with the example but is incomplete. it can be removed, use cases are good examples}
%However, the formalization does not yet specify \emph{how} the system achieves the quality attribute outcome, that is, what concrete mechanism is applied.
%%This is the gap we address.
%\subsection{Extended Formalization: Adding Tactics}
%We extend the formalization by introducing \emph{architectural tactics} as an additional input.
The main aspect being the fact that the \textbf{same uncertainty interaction} (same $IC, Types, Design, Cr, Q$) can yield \textbf{different quality attribute outcomes} depending on which tactic is applied.
This makes the tactics the \emph{degree of freedom} available to the architect for regulating quality attribute responses under uncertainty. This formalization serves as a logical framework that maps identified uncertainty interactions to the most effective mitigating tactics and their associated trade-offs.

\subsection{Trade-off Characterization}

A critical property of tactics is that they involve trade-offs.
Applying a tactic may assure certain quality attributes ($\mathbf{Q}^{+}$) while potentially degrading others ($\mathbf{Q}^{-}$).
We capture this by distinguishing:

\begin{equation}
\label{eq:tradeoff}
\begin{aligned}
\mathbf{Q}^{+} &= \{q_j \mid \text{tactic assures or improves } q_j\} \\
\mathbf{Q}^{-} &= \{q_j \mid \text{tactic may degrade } q_j\}
\end{aligned}
\end{equation}

\noindent The \emph{trade-off profile} of a tactic is the pair $(\mathbf{Q}^{+}, \mathbf{Q}^{-})$.
For two tactics $T_1, T_2$ applicable to the same uncertainty interaction, comparing $(\mathbf{Q}^{+}_1, \mathbf{Q}^{-}_1)$ against $(\mathbf{Q}^{+}_2, \mathbf{Q}^{-}_2)$ enables principled architectural decision-making. 

%\todo{This requires additional analysis}

\subsection{Composition of Tactics}

In practice, multiple tactics are often composed to address an uncertainty interaction.
Given tactics $T_1, \ldots, T_k$, the composed output satisfies:
\begin{equation}
\label{eq:composition}
\mathbf{Q}^{+}(T_1 \circ \cdots \circ T_k) \supseteq \bigcup_{i=1}^{k} \mathbf{Q}^{+}(T_i)
\end{equation}
where the superset relationship reflects that composed tactics may achieve quality attributes that no individual tactic achieves alone (emergent assurance).
However, composition may also introduce new conflicts:
\begin{equation}
\label{eq:conflict}
\mathbf{Q}^{-}(T_1 \circ \cdots \circ T_k) \not\subseteq \bigcup_{i=1}^{k} \mathbf{Q}^{-}(T_i)
\end{equation}
meaning that combining tactics can create degradation not present in any individual tactic.

%% ============================================================
\section{Use Cases and Tactic Exemplification}
\label{sec:usecases}

\subsection{Use Case 1: SLO Compliance in LLM Serving System}
\label{sec:uc1}

\textbf{Context (Managed System).}
Modern LLM serving faces two dominant uncertainties that inflate tail latency. $U_1$: workload uncertainty (requests vary widely in prompt and completion length, shifting the KV-cache working set and the batching/parallelism trade-off); and $U_2$: model/output uncertainty (stochastic decoding yields variable output lengths even for similar prompts, making the decode horizon only partially knowable at admission time).
At decode, each active sequence incurs KV reads/writes per step; as responses lengthen, steps become effectively memory-bandwidth bound, so widening batches (higher parallelism) can slow per-step progress and worsen p95/p99 (the latency threshold below which 95\%/99\% of requests finish), despite high average utilization.
These aspects compound in agentic AI where multiple LLMs are combined to achieve a common goal.

\textbf{Uncertainty Interaction.}
Workload and output uncertainties exhibit \emph{augmentation}: the aleatory variability in requests amplifies the effect of  output lengths, jointly inflating tail latency in a way that neither source alone would cause.

\textbf{Formalization Instance.}

\smallskip
\noindent\hspace{1em}%
\texttt{Input:}\; $\big($%
IC: $\langle$Augmentation$\rangle$,\;
Types: [ Aleatory ],

\noindent\hspace{4.2em}%
Design: $\langle$ DT: Neural network and queueing model,

\noindent\hspace{5.8em}%
CV: [ temp, top\_p ] $\rangle$,\;
Cr: $[]$,

\noindent\hspace{4.2em}%
$\mathbf{Q}$: [ performance, throughput, cost efficiency, 

\noindent\hspace{6em}%
tail latency, scheduling accuracy, 

\noindent\hspace{6em}%
output quality variance, prediction overhead,

\noindent\hspace{6em}%
memory contention],

\noindent\hspace{4.2em}%
Tactics: $[ \langle$see below$\rangle ] \big)$

\noindent\hspace{1em}%
\texttt{Output:}\; $\mathbf{Q}^{+}\langle$ performance, $T_{1.2}\rangle$

\smallskip
\textbf{Tactics Catalog for UC1:}

\smallskip
\noindent\emph{$T_{1.1}$ -- Batching of Prompts} $\langle$prompt aggregation; system; runtime$\rangle$.
Groups incoming prompts into batches to amortize GPU kernel launch overhead and improve throughput, at the risk of increasing per-request latency when batch sizes grow too large.

\noindent\hspace{1em}\texttt{Output:}\; $\mathbf{Q}^{+}\langle$throughput, GPU utilization$\rangle$

\noindent\hspace{1em}$\mathbf{Q}^{-}$: $\langle$tail latency (p95/p99), fairness for short requests$\rangle$.

\smallskip
\noindent\emph{$T_{1.2}$ -- Routing of Prompts to Different LLMs} (model-aware routing; system; runtime).
Routes requests to different LLM instances (or model sizes) based on estimated complexity, balancing cost against quality.

\noindent\hspace{1em}\texttt{Output:}\; $\mathbf{Q}^{+}\langle$cost efficiency, latency balance$\rangle$

\noindent\hspace{1em}$\mathbf{Q}^{-}$: $\langle$output quality variance, routing overhead$\rangle$.

\smallskip
\noindent\emph{$T_{1.3}$ -- Prediction of Response Length} $\langle$output length estimation; component; runtime$\rangle$.
Predicts the expected response length before decode begins, enabling informed scheduling and resource allocation decisions.

\noindent\hspace{1em}\texttt{Output:}\; $\mathbf{Q}^{+}\langle$scheduling accuracy, resource predictability$\rangle$

\noindent\hspace{1em}$\mathbf{Q}^{-}$: $\langle$prediction overhead, misprediction cost$\rangle$.

\smallskip
\noindent\emph{$T_{1.4}$ -- Scheduling of Parallel Batches} $\langle$parallel batch orchestration; system; runtime$\rangle$.
Schedules multiple batches in parallel across GPU resources, balancing memory-bandwidth utilization against per-step decode progress.

\noindent\hspace{1em}\texttt{Output:}\; $\mathbf{Q}^{+}\langle$throughput, SLO compliance rate$\rangle$

\noindent\hspace{1em}$\mathbf{Q}^{-}$: $\langle$memory contention, implementation complexity$\rangle$.

\subsection{Use Case 2: Agentic System for Autonomous CloudOps}
\label{sec:uc2}

\textbf{Context (Managed System).}
CloudOps manages cloud infrastructure through automated workflows.
Organizations navigating complex cloud environments require autonomous systems capable of executing diverse operational tasks at scale.
CloudOps platforms are increasingly leveraging \ac{AI} agents to automate cloud management tasks, including resource provisioning, compliance enforcement, and incident response~\cite{parthasarathy2025cloudops}.
Two critical uncertainty sources include $U_1$: memory staleness and retrieval gaps (the stored context may be outdated and retrieval may be incomplete, low recall, even when precise); and $U_2$: context window limitation (the bounded context window of LLM-based agents induces partial observability of the system state and prior reasoning steps).

\textbf{Uncertainty Interaction.}
$U_1$ and $U_2$ exhibit \emph{dominance} ($U_1 > U_2$): memory staleness dominates context window limitations because even when the context window is sufficient, stale or incomplete memory renders the available context unreliable.
The epistemic uncertainty in stored beliefs compounds with aleatory variability in retrieval quality.

\textbf{Formalization Instance.}

\smallskip
\noindent\hspace{1em}%
\texttt{Input:}\; $\big($%
IC: $\langle$Dominance -- $U_1 > U_2\rangle$,

\noindent\hspace{4.2em}%
Types: [ Epistemic, Aleatory ],

\noindent\hspace{4.2em}%
Design: $\langle$ DT: $S$, CV: [ temperature, top\_p ] $\rangle$,\;
Cr: $[U_1]$,

\noindent\hspace{4.2em}%
$\mathbf{Q}$: [ memory freshness, context completeness, 

\noindent\hspace{6em}%
context coverage, information loss, 

\noindent\hspace{6em}%
functional correctness ],

\noindent\hspace{4.2em}%
Tactics: $ [ \langle$see below$\rangle ] \big)$ 

\noindent\hspace{1em}%
\texttt{Output:}\; $\mathbf{Q}^{+}\langle$user satisfaction, $T_{2.3}\rangle$

\smallskip
\noindent where $S$ denotes the overall system design of the multi-agent CloudOps platform.

\smallskip
\textbf{Tactics Catalog for UC2:}

\smallskip
\noindent\emph{$T_{2.1}$ -- Memory Staleness Mitigation} $\langle$drift-aware retrieval; component; runtime$\rangle$.
Monitors the drift of stored metrics (\eg recall degradation over time) and triggers re-indexing or cache invalidation when staleness exceeds thresholds.

\noindent\hspace{1em}\texttt{Output:}\; $\mathbf{Q}^{+}\langle$memory freshness, retrieval recall$\rangle$

\noindent\hspace{1em}$\mathbf{Q}^{-}$: $\langle$re-indexing cost, retrieval latency$\rangle$.

\smallskip
\noindent\emph{$T_{2.2}$ -- Knowledge Augmentation} $\langle$external knowledge injection; subsystem; runtime$\rangle$.
Supplements agent memory with up-to-date external knowledge sources (\eg documentation, API changelogs, compliance databases) to compensate for retrieval gaps.

\noindent\hspace{1em}\texttt{Output:}\; $\mathbf{Q^{+}}\langle$context completeness, decision accuracy$\rangle$

\noindent\hspace{1em}$\mathbf{Q}^{-}$: $\langle$knowledge integration overhead, potential noise$\rangle$.

\smallskip
\noindent\emph{$T_{2.3}$ -- Hierarchical Summarization} $\langle$multi-level context compression; component; runtime$\rangle$.
Applies hierarchical summarization to long conversations and operational histories, fitting more relevant context within the bounded context window while preserving critical information.

\noindent\hspace{1em}\texttt{Output:}\; $\mathbf{Q}^{+}\langle$context coverage, reasoning coherence$\rangle$

\noindent\hspace{1em}$\mathbf{Q}^{-}$: $\langle$information loss, summarization accuracy$\rangle$.

\subsection{Use Case 3: Proactive Continuous Operations Using LLMs and AIOps}
\label{sec:uc3}

\textbf{Context (Managing System).}
In this case, Digital Twins (DTs) are used as accurate virtual replicas of production software applications running on IT environments (Real Twins) to achieve proactive failure avoidance.
By leveraging DT-driven ``what-if'' simulations and experimentation, the system evaluates the risk of operational changes (\eg configuration updates) before they reach production.
\ac{AI} and \acp{LLM} are integrated to automate scenario-based experimentation and simulation, synthesize insights from gathered and analyzed data, and provide human-readable explanations for identified risks.
In this use case, critical uncertainties include $U_1$: the epistemic uncertainty introduced by the LLM model itself (what the model captures about the system); and $U_2$: the aleatory nature of operational data and metrics (noisy, variable sensor readings, the dynamic nature of hardware resources use by software application in different execution scenarios, and, in general, context monitoring issues in telemetry) that are used to continuously re-train the models.

\textbf{Uncertainty Interaction.}
Model uncertainty and data uncertainty exhibit \emph{dominance} ($U_1 > U_2$): the epistemic uncertainty in the \acp{LLM} learned representations dominates the aleatory variability in the operational data gathered through monitoring, because model errors systematically bias predictions regardless of data variability and quality.

\textbf{Formalization Instance.}

\smallskip
\noindent\hspace{1em}%
\texttt{Input:}\; $\big($%
IC: $\langle$Dominance$\rangle$,\;
Types: [ Epistemic, Aleatory ],

\noindent\hspace{4.2em}%
Design: $\langle$DT: $LLM$ pre-trained and then re-trained,

\noindent\hspace{7.8em}%
CV: [ diagnostic error, resource use,

\noindent\hspace{7.8em}%
mispredictions ] $\rangle$,\;

\noindent\hspace{7.8em}%
Cr: $[U_1]$,

\noindent\hspace{4.2em}%
$\mathbf{Q}$: [ diagnostic accuracy, distribution shift risk,

\noindent\hspace{6em}%
prediction robustness, inference cost, 

\noindent\hspace{6em}%
 latency, performance ],

\noindent\hspace{4.2em}%
Tactics: $ [\langle$see below$\rangle ] \big)$ 

\noindent\hspace{1em}%
\texttt{Output:}\; $\mathbf{Q}^{+}\langle$prediction robustness, $T_{3.2}\rangle$

\smallskip
\textbf{Tactics Catalog for UC3:}

\smallskip
\noindent\emph{$T_{3.1}$ -- Generation of Synthetic Data} $\langle$labeled data synthesis; subsystem; design-time$\rangle$.
Generates synthetic operational data labeled with root causes and fix recommendations, expanding the training distribution to reduce epistemic uncertainty in the model's understanding of failure modes.

\noindent\hspace{1em}\texttt{Output:}\; $\mathbf{Q}^{+}\langle$diagnostic accuracy, failure coverage$\rangle$

\noindent\hspace{1em}$\mathbf{Q}^{-}$: $\langle$distribution shift risk, generation cost$\rangle$.

\smallskip
\noindent\emph{$T_{3.2}$ -- Ensembles} $\langle$multi-model aggregation; system; runtime$\rangle$.
Deploys multiple models (or multiple LLM configurations) and aggregates their predictions, reducing the impact of any single model's epistemic errors through diversity.

\noindent\hspace{1em}\texttt{Output:}\; $\mathbf{Q}^{+}\langle$prediction robustness, model ensemble$\rangle$

\noindent\hspace{1em}$\mathbf{Q}^{-}$: $\langle$inference cost, latency, resource consumption$\rangle$.

\smallskip
\noindent\emph{$T_{3.3}$ -- Knowledge Augmentation} $\langle$external knowledge injection; subsystem; runtime$\rangle$.
Supplements LLM with knowledge hyper-graphs to compensate for imbalanced data on relationships between components and failure-sources.

\noindent\hspace{1em}\texttt{Output:}\; $\mathbf{Q^{+}}\langle$context completeness, decision accuracy$\rangle$

\noindent\hspace{1em}$\mathbf{Q}^{-}$: $\langle$knowledge consistency, failure coverage$\rangle$.

%% ============================================================
\section{Cross-Cutting Analysis}
\label{sec:analysis}

\subsection{Tactic Patterns Across Use Cases}

Table~\ref{tab:tactic_patterns} summarizes the tactic-to-quality-attribute mappings across all three use cases.
The table makes explicit the relationship between uncertainty interaction categories, tactics, and quality attribute trade-offs.

\begin{table*}[!t]
  \centering
  \caption{Summary of Uncertainty Interactions, Tactics, and Quality Attribute Trade-offs Across Use Cases}
  \label{tab:tactic_patterns}
  \renewcommand{\arraystretch}{1.2}
%\footnotesize
\begin{tabular}{@{}l l l l l l l@{}}
\toprule
\textbf{UC} & \textbf{IC} & \textbf{Tactic} & \makecell[l]{\textbf{Scope /}\\\textbf{Lifecycle}} & \textbf{Cr} & \textbf{$\mathbf{Q}^{+}$ (Assured)} & \textbf{$\mathbf{Q}^-$ (At Risk)} \\
\midrule
\multirow{4}{*}{\rotatebox[origin=c]{90}{\scriptsize\textbf{UC1}}}
 & \multirow{4}{*}{\makecell[l]{Augmentation}}
 & Batching of Prompts       & Sys / RT    & $[\,]$          & Throughput, GPU util.            & Tail lat., fairness (short) \\
 & & Routing to Diff.\ LLMs   & Sys / RT    & $[\,]$          & Cost efficiency, lat.\ balance   & Output quality var., overhead \\
 & & Resp.\ Length Prediction  & Comp / RT   & $[\,]$          & Sched.\ accuracy, predictability & Prediction overhead, mispredict \\
 & & Parallel Batch Sched.    & Sys / RT    & $[\,]$          & Throughput, SLO compliance       & Memory contention, complexity \\
\midrule
\multirow{3}{*}{\rotatebox[origin=c]{90}{\scriptsize\textbf{UC2}}}
 & \multirow{3}{*}{\makecell[l]{Dominance}}
 & Memory Staleness Mit.     & Comp / RT   & $[U_1]$         & Memory freshness, recall         & Re-index cost, retrieval lat. \\
 & & Knowledge Augment.       & Sub / RT    & $[U_1]$         & Context completeness, accuracy   & Integration overhead, noise \\
 & & Hierarchical Summ.       & Comp / RT   & $[U_1]$         & Context coverage, coherence      & Information loss, summ.\ accuracy \\
\midrule
\multirow{2}{*}{\rotatebox[origin=c]{90}{\scriptsize\textbf{UC3}}}
 & \multirow{2}{*}{\makecell[l]{Dominance}}
 & Synthetic Data Gen.       & Sub / DT    & $[U_1]$         & Diagnostic accuracy, coverage    & Distrib.\ shift, gen.\ cost \\
 & & Ensembles                & Sys / RT    & $[U_1]$         & Pred.\ robustness, func.\ corr.  & Inference cost, lat., resources \\
 & & Knowledge Augment.       & Sub / RT    & $[U_1]$         & Context completeness, accuracy   & Knowledge consistency, failure cover. \\
\bottomrule
\end{tabular}
\\[4pt]
\raggedright\scriptsize
\textbf{Legend:} RT\,:\,Run-time, DT\,:\,Design-time, Sys\,:\,System, Comp\,:\,Component, Sub\,:\,Subsystem. UC1\,:\,SLO Compliance in LLM Serving, UC2\,:\,Agentic CloudOps, UC3\,:\,AIOps with Digital Twins.
\end{table*}

Several patterns emerge from this analysis:

\textbf{Quality of Service and Cost Trade-off.}
Across all three use cases, tactics that improve correctness, accuracy, performance, or reliability tend to introduce additional cost, which can be computational (ensembles in UC3), temporal (re-indexing in UC2), or in terms of system complexity (routing in UC1).
The tactic selection therefore, involves explicit cost-quality
trade-offs that the formalization makes transparent through
($Q^+, Q\text{--}$).

\textbf{Interaction Category Shapes Tactic Selection.}
The uncertainty interaction category (IC) has a direct influence on tactic design.
For \emph{augmentation} interactions (UC1), tactics focus on \emph{reducing} the amplification, \eg batching and parallel scheduling prevent workload variability from amplifying output variability.
For \emph{dominance} interactions (UC2, UC3), tactics target the \emph{dominant} source directly, \eg memory staleness mitigation addresses $U_5$ which dominates $U_7$, and synthetic data generation addresses the epistemic model uncertainty ($U_1$) that dominates the aleatory data uncertainty.

\textbf{Managed vs.\ Managing System Distinction.}
UC1 and UC2 treat the LLM-based system as the \emph{managed system} whose uncertainty must be controlled, while UC3 treats the LLM as the \emph{managing system} that must itself be made more reliable.
This distinction influences where tactics are applied: managed-system tactics operate on the system's inputs and runtime behavior, while managing-system tactics improve the model's knowledge and robustness.

\textbf{Runtime Dominance.}
The majority of tactics operate at runtime, reflecting that \ac{GAI} uncertainty is largely irreducible at design time due to the stochastic nature of LLM outputs.
The notable exception is synthetic data generation (UC3), which is a design-time tactic that reduces epistemic uncertainty by expanding the model's training distribution.

%
%\textbf{}{Cost of Assurance.}
%Every tactic that provides quality attribute assurance introduces cost.
%Ensembles (UC3) multiply inference cost by the number of models.
%Hierarchical summarization (UC2) risks information loss.
%Response length prediction (UC1) adds a prediction step before each request.
%The tactic selection therefore involves explicit cost-quality trade-offs that the formalization makes transparent through $(\mathbf{Q}^{+}, \mathbf{Q}^{-})$.
%
\subsection{Interaction Between Tactics}

When tactics are composed, interactions can be synergistic or conflicting:

\textbf{Synergy.}
In UC1, combining response length prediction ($T_{1.3}$) with parallel batch scheduling ($T_{1.4}$) achieves better SLO compliance than either alone, because predicted lengths inform the scheduler's batch composition decisions.

\textbf{Conflict.}
In UC2, knowledge augmentation ($T_{2.2}$) may conflict with hierarchical summarization ($T_{2.3}$): adding external knowledge expands the context that then needs to be compressed, potentially causing the summarization to discard the very knowledge that was just injected.

\textbf{Enablement.}
In UC3, synthetic data generation ($T_{3.1}$) enables ensembles ($T_{3.2}$) by providing sufficient labeled examples to train diverse ensemble members.
Without adequate labeled data, ensemble diversity is limited and the robustness benefit is reduced.

%% ============================================================
\section{Towards Mitigation Patterns}
\label{sec:patterns}

We envision the tactic catalog evolving into a catalog of \emph{uncertainty mitigation patterns}, reusable combinations of uncertainty interactions, tactics, and expected quality attribute outcomes.
A pattern captures: 
\begin{enumerate*}[label=(\arabic*)]
\item the interaction category and uncertainty types that motivate it, 
\item the system design contexts in which it is applicable, 
\item the tactics it prescribes, and 
\item the quality attribute trade-off profile $(\mathbf{Q}^{+}, \mathbf{Q}^{-})$.
\end{enumerate*}

For instance, a ``\emph{Predictive Batch Orchestration}'' pattern for LLM serving (IC: Augmentation) would prescribe composing response length prediction, batching, and parallel scheduling, assuring $\mathbf{Q}\langle$performance, cost$\rangle$ with controlled scheduling overhead.
A ``\emph{Memory-Aware Agent}'' pattern for CloudOps (IC: Dominance) would compose memory staleness mitigation, knowledge augmentation, and hierarchical summarization, assuring $\mathbf{Q}\langle$user satisfaction$\rangle$ while managing information loss.
A ``\emph{Robust AIOps Predictor}'' pattern (IC: Dominance) would compose synthetic data generation with ensembles, assuring $\mathbf{Q}\langle$functional correctness$\rangle$ at higher inference cost. This pattern-level thinking is critical because individual tactics, while useful, do not fully address the compounding nature of \ac{GAI} uncertainty.
It is the \emph{composition} of tactics and understanding their interactions that enables effective mitigation.

%% ============================================================
\section{Vision and Open Challenges}
\label{sec:challenges-agenda}

\noindent \textbf{CH1: Uncertainty Interaction Characterization in Gen-AI Based Systems}. A foundational challenge is the lack of systematic methods for characterizing uncertainty interactions in Gen-AI based systems. While prior work has established general uncertainty dimensions and interaction categories, Gen-AI systems introduce novel uncertainty sources, such as hallucinations, stochastic output variability, memory staleness, and tool orchestration uncertainty, that interact across components, agents, and layers of the architecture. These interactions are often implicit, emergent, and only observable through their downstream effects on system qualities. The vision is to move from ad hoc identification of uncertainty interactions toward explicit, architecture-aware representations that make such interactions visible, comparable, and analyzable. Achieving this requires extending existing uncertainty frameworks with Gen-AI–specific concepts and grounding interaction characterization in architectural abstractions that reflect how modern Gen-AI systems are actually built and operated.

% \smallskip
\noindent \textbf{CH2: Guidance for Selecting Mitigation Strategies}. Even when uncertainty interactions are recognized, system architects currently lack practical guidance on how to mitigate them. The landscape of possible mitigation strategies is broad, spanning model-level techniques, architectural tactics, and operational interventions. Without systematic guidance, mitigation decisions are often based on experience, trial-and-error, or isolated best practices, limiting their scalability and transparency. The vision here is that mitigation decisions should be driven by a clear understanding of the underlying uncertainty interactions and their architectural context. The mitigation function provides a conceptual foundation for this reasoning, but it must evolve into a form that supports principled selection and comparison of mitigation tactics. This includes aligning interaction categories with families of tactics, considering system criticality, and making explicit the anticipated impact on quality attributes.

\noindent \textbf{CH3: Architectural Complexity and Cross-Layer Uncertainty Interactions}. A defining characteristic of Gen-AI based systems is their architectural complexity. These systems frequently integrate multiple models, external services, data sources, and agentic components through layered and highly dynamic architectures. In such settings, uncertainty rarely remains confined to a single component; instead, it propagates across layers and subsystems, sometimes amplifying or transforming in the process. This challenge calls for an architectural perspective on uncertainty mitigation that goes beyond local fixes. The vision is to identify architectural patterns and design strategies that either constrain uncertainty propagation or provide systematic points of control where mitigation can be applied effectively. Such strategies must account for uncertainty across design-time, deployment-time, and runtime, reflecting the fact that Gen-AI uncertainty is largely irreducible and must be continuously managed rather than eliminated.

\noindent \textbf{CH4: Trade-Off Evaluation of Mitigation Strategies}. Mitigation strategies inevitably introduce trade-offs. Techniques that improve reliability, safety, or correctness may negatively impact performance, cost, or usability, and these trade-offs are often amplified in Gen-AI systems due to the computational and operational costs of AI components. Despite this, there is currently limited systematic support for evaluating mitigation strategies in terms of their quality attribute impact, particularly under interacting uncertainties. The vision is to make trade-offs an explicit and analyzable part of mitigation reasoning. This requires evaluation approaches that move beyond task-centric metrics and instead assess how mitigation strategies affect broader system qualities in realistic operational conditions. Making such trade-offs visible is essential for informed architectural decision-making in safety-, business-, or mission-critical Gen-AI systems.  

\noindent \textbf{CH5: Dynamic Context Management}. Integrating GenAI into software-intensive systems poses significant dynamic context management challenges rooted in epistemic stochasticity, where non-deterministic outputs and hallucinations prevent the exhaustive design-time specification of monitoring requirements and relevant context sources. A primary challenge is maintaining context relevance, as the stochastic nature of GenAI demands that monitoring infrastructures self-adapt at runtime to assure the validity of context used to reason. This is exacerbated by memory staleness, which acts as a dominant uncertainty over context window limitations; if retrieved context is outdated or irrelevant, even an expansive window becomes useless. Consequently, the context information lifecycle must be managed at runtime to assure the relevance of context variables that must be monitored as new ones may appear or disappear due to uncertainty interactions, and to maintain the pertinence of monitoring strategies~\cite{villegas2013dynamico}.

\noindent \textbf{Roadmap Towards Systematic Mitigation}. Addressing these challenges requires a staged research roadmap. In the near term, efforts should focus on consolidating conceptual foundations for characterizing uncertainty interactions in Gen-AI systems, including refined taxonomies and architecture-aware representations. These foundations enable early forms of mitigation guidance by organizing existing tactics in relation to interaction categories and architectural scope. In the mid term, research should emphasize architectural integration, focusing on uncertainty-aware design strategies, mitigation patterns, and support for reasoning about tactic composition in complex systems. At this stage, mitigation becomes part of architectural thinking rather than an after-the-fact correction mechanism. In the long term, the emphasis shifts toward rigorous evaluation and adaptive mitigation. Systematic evaluation frameworks are needed to empirically assess trade-offs and to compare alternative mitigation approaches. Building on these foundations, Gen-AI systems can increasingly incorporate adaptive and partially autonomous mitigation mechanisms that respond to emerging uncertainty interactions at runtime while remaining transparent and controllable.

\section{Related Work}
\label{sec:related}

Our framework builds on the body of work addressing uncertainty in self-adaptive software systems. Early work focuses on defining uncertainty, identifying its sources, classifying uncertainties and structuring taxonomies~\cite{ramirez2012taxonomy,esfahani2013uncertainty,mahdavi2017classification}. Such definitions provide a foundation for proposing approaches and defining a research agenda for addressing self-adaptation under uncertainty~\cite{weyns2023b}. However, with further proliferation of ML/AI techniques to effect self-adaptation, software systems have grown increasingly complex, where individual sources of uncertainty and their combinations and interactions need to be managed and mitigated~\cite{shonan232}. \acp{LLM} are becoming integral to various software systems, introducing new and additional sources of uncertainty~\cite{jialong2024}. In GenAI-based software systems, which we explicitly address in our case studies, inherent \ac{AI} model uncertainties are identified as one of the primary challenges in performance evaluation~\cite{akshathala2025beyond}. 
Our proposal takes the field a step further by proposing the framework for mitigation strategies to manage these interactions between multiple sources of uncertainty.

%% ============================================================
\section{Conclusion}
\label{sec:conclusion}

\ac{GAI}-based software intensive systems introduce qualitatively new forms of uncertainty and their interactions that demand novel mitigation approaches and techniques. This vision paper proposes an uncertainty interaction formalization that introduces architectural tactics as an explicit input, making the mapping from uncertainty interactions to quality attribute outcomes actionable.
%The key insight is that the uncertainty interaction category (\ie dominance, augmentation, conflict, and independence) shapes which tactics are appropriate, and the tactic is the degree of freedom through which architects influence quality attribute responses.
By grounding this formalization in three use cases: LLM serving, AIOps with Software Twins, and agentic CloudOps,  we demonstrate that the framework enables principled reasoning about trade-offs, tactic composition, and mitigation patterns. Further, we present vision,  open challenges and a roadmap to the research community towards mitigating uncertainty interactions in GenAI-based systems.
%Such a framework, and the accompanying research agenda can bridge the gap between the 
%self-adaptive systems community's mature uncertainty interaction frameworks and the practical challenges of deploying \ac{GAI} systems at scale.

\bibliographystyle{IEEEtran}

\end{document}